\documentstyle[preprint,aps]{revtex}

\tighten
\begin{document}
\draft

\preprint{\vbox{\baselineskip=12pt
\rightline{gr-qc/9412070}
\rightline{Accepted for Publication in Physical Review D}}}

\title{Numerical Black Holes: A Moving Grid Approach}

\author{C. Bona}
\address{Departament de F\'\i sica. Universitat de les Illes Balears.  \\
   E-07071 Palma de Mallorca, SPAIN.}

\author{J. Mass\'o\thanks{On leave of absence from Departament de F\'\i sica.
Universitat de les Illes Balears. E-07071 Palma de Mallorca, SPAIN.}}
\address{National Center for Supercomputer Applications. \\
605 E. Springfield Avenue, Champaign-Urbana. IL 61821, USA.}

\author{J. Stela}
\address{Departament de F\'\i sica. Universitat de les Illes Balears.  \\
   E-07071 Palma de Mallorca, SPAIN.}

\maketitle
\begin{abstract}

Spherically symmetric (1D) black-hole spacetimes are considered as a
test for numerical relativity. A finite difference code, based in the
hyperbolic structure of Einstein's equations with the harmonic slicing
condition is presented. Significant errors in the mass function are
shown to arise from the steep gradient zone behind the black hole
horizon, which challenge the Computational Fluid Dynamics numerical
methods used in the code.  The formalism is extended to moving
numerical grids, which are adapted to follow horizon motion.  The
black hole exterior region can then be modeled with higher accuracy.

\end{abstract}

\narrowtext

\section{INTRODUCTION.}

Black hole simulations are known to provide a severe test of Numerical
Relativity, even in the one-dimensional (1D) case\cite{Hobill}. These
strong field scenarios imply a wide dynamical range involving very
different time and length scales.  The coordinate degrees of freedom
must be used with special care in order to prevent the numerical code
to crash when spacetime approaches a singularity. This is the idea
underlying the different singularity avoiding conditions which have
been proposed and tested in Numerical
Relativity\cite{Smarr,Piran,PR88}. This approach, however, is not free
from problems. Spacetime dynamics gets locally frozen near the
singularity, leading to an abrupt transition zone around the horizon.
This can cause either steep space gradients\cite{Shapiro} or spikes in
the radial metric function\cite{Smarr} which make difficult to
maintain the accuracy or even lead to code crashing during numerical
evolution.

A major step towards a singularity-proof scheme in Numerical
Relativity is the use of a horizon boundary condition. The
idea\cite{Unruh,Thornburg} is to reduce drastically the dynamical
range by evolving just the observable region and imposing a suitable
boundary condition on or slightly inside the horizon, which is a
one-way membrane. This idea has been actually implemented in
Ref.~\cite{Seidel} by the combined use of a ``horizon locking''
coordinate system and a ``causal'' finite difference scheme which is
very similar to the ``causal reconnection'' scheme introduced in
Ref.~\cite{Alcubierre}.  Numerical evolution was there shown to
proceed without significant errors beyond the limit of $t=100\:m$,
where the previous codes that used fixed boundaries \cite{Hobill}
became unstable or extremely inaccurate.  The results presented in
Ref.~\cite{Hobill} correspond to the maximal slicing
condition\cite{Smarr}. This does not mean that that code could not use
alternative slicings, like the harmonic one. It is simply due to the
fact that maximal slicing happens to be very robust, leading to longer
numerical black hole evolution than other singularity avoiding
slicings\cite{Hobill}.

Singularity avoidance, however, is not our only guideline in
constructing a Numerical Relativity code. We want to use a system of
evolution equations which actually translates the causal structure of
the spacetime: it should be a hyperbolic system of partial
differential equations with the local speed of light as characteristic
speed. It has been shown recently\cite{PRL92,Report} that the use of a
harmonic time coordinate (harmonic slicing\cite{PR88}) leads to such
hyperbolic evolution systems, which can also be expressed in
flux-conservative form. In that way, we can get the well known
structure of the hydrodynamic equations and the huge arsenal of the
Computational Fluid Dynamics (CFD) methods is at our disposal. The
straightforward use of such methods has yet resulted into a 3D
numerical code\cite{Southampton} which is able by now to evolve vacuum
spacetimes admitting periodic boundary conditions.

 In the present work, we will use CFD methods to improve the quality
of harmonic slicing black hole codes.  The first part (sections 2-4)
of the paper deals with the standard approach, where we use a finite
difference discretization in a fixed grid. The horizon boundary
condition is implemented in the second part (sections 5-8) by using a
moving numerical grid.  The use of a moving grid is not new in
Numerical Relativity.  Wilson\cite{Wilson} used it to deal with the
Relativistic Hydrodynamics equations. The same ''adaptive mesh''
technique was then extended to the field equations and successfully
applied to the study of axisymmetric stellar
collapse\cite{Dykema,Evans}. Here we provide a new application of this
technique to the study of black hole evolution, where we use the grid
speed to keep the numerical mesh outside the horizon. This approach is
compared with the ``horizon locking'' condition introduced in
Ref.~\cite{Seidel}.

\section{THE 1D BLACK HOLE.}
Let us write the spherically symmetric (1D) vacuum line element in
isotropic coordinates:
\begin{equation}
ds^2 = - \left(\frac{\rho-m/2}{\rho+m/2}\right)^2\:dt^2 +
       \left(1+\frac{m}{2\rho}\right)^4\:(d\rho^2+\rho^2\:d\Omega),
       \label{static}
\end{equation}
which is locally isometric to the Schwarzschild metric. The lapse function
\begin{equation}
\alpha = \frac{\rho -m/2}{\rho +m/2}                           \label{lapse}
\end{equation}
vanishes at $\rho=m/2$, which corresponds to the black hole horizon ($r=2m$ in
standard Schwarzschild coordinates).

The form (\ref{static}) of the line element is convenient for
numerical applications: (i) it is easy to express
(\ref{static}) in Cartesian coordinates; this allows one to consider
the same problem as a test for generic (3D) numerical codes,
and (ii) the metric (\ref{static}) is invariant under the inversion
symmetry
\begin{equation}
\frac{\rho}{m/2} \rightarrow \frac{m/2}{\rho}              \label{inversion}
\end{equation}
mapping the black hole outer region ($\rho>m/2$) into the inner part
and vice versa; this provides a consistent inner boundary condition at
the inversion point $\rho=m/2$ (the ``throat''), allowing one to
evolve just the exterior part of the black hole spacetime
(``wormhole'' evolution: see Ref. \cite{Hobill}).

We shall use the space part of the line element (\ref{static}) to
provide the initial data for our numerical models. The singular lapse
function (\ref{lapse}) will be however replaced by a non-singular
symmetric function with initial value
\begin{equation}
\alpha(0,\rho) = constant					\label{lapse1}
\end{equation}
in order to get a different spacetime slicing so that the metric
components are no longer static but evolving in time. The line
element is expressed in the generic spherically symmetric form
\begin{equation}
ds^2 = -\alpha^2(t,\rho)\:dt^2 + X^2(t,\rho)\:d\rho^2
       + Y^2(t,\rho)\:d\Omega.	  				\label{metric}
\end{equation}

The initial condition (\ref{lapse1}) preserves the inversion symmetry
(\ref{inversion}) of the spacetime. This means that the throat
connecting the two isometric sheets will remain fixed at $\rho=m/2$
and inner boundary conditions can be imposed there consistently at
every step of the evolution process. Note however that the black hole
horizon\cite{apparent} moves with the local light speed
\begin{equation}
c = \alpha/X,							\label{light}
\end{equation}
which will be now different from zero everywhere. This means that the
horizon will start moving outwards, away from the fixed
throat position.  The coincidence between throat and horizon in the
static form (\ref{static}) was because the singular lapse choice
(\ref{lapse}) caused the vanishing of the local speed of light at that
point.

In order to locate the black hole horizon\cite{apparent}, we will make use
of the apparent horizon local definition
\begin{equation}
\dot{Y}/\alpha + Y^\prime /X = 0,			\label{horizon}
\end{equation}
where dots and primes stand for time and space derivatives,
respectively. The mass function $M(t,\rho)$, defined by
\begin{equation}
2M/Y = 1 + [(\dot{Y}/\alpha)^2-(Y^\prime /X)^2],		\label{Bondi}
\end{equation}
gives at every instant the total amount of mass enclosed by a sphere of
coordinate radius $\rho$. Of course, it
does coincide with the constant parameter $m$ in our case. Condition
(\ref{horizon}) then implies that $Y=2m$ at the black hole horizon.

\section{THE NUMERICAL CODE.}
Our code is a finite difference version of Einstein field equations in
first order form. This means that the set of basic quantities includes
not only the lapse function $\alpha$ and the spatial metric components
(the contravariant ones $g^{ij}$ in our case), but also their first
order space and time derivatives. The time derivative of the lapse
function is given by imposing the time coordinate to be harmonic
(harmonic slicing \cite{PR88}). For the sake of simplicity, the
remaining set of independent quantities will be described in what
follows as the components of a single vector valued function ${\bbox
u}$. These quantities are taken to be independent because we consider
the constraint equations as first integrals of our evolution system,
which are imposed on the initial data only (free evolution
approach). The conservation of the constraints can then be used as an
accuracy test.

Under these conditions, it has been shown \cite{PRL92} that in the
generic 3D case the evolution system can be written as an hyperbolic
system of balance laws
\begin{equation}
\partial_t {\bbox u} + \partial_k {\bbox F}^k({\bbox u})
      = {\bbox S}({\bbox u}),
\label{balance} \end{equation}
where the fluxes ${\bbox F}^k$ and sources ${\bbox S}$ are vector
valued functions of ${\bbox u}$. Moreover, it has been shown
\cite{Report} that there is an infinite family of hyperbolic evolution
systems which share only the physical solutions (the ones actually
satisfying the constraint equations). We will use here a spherically
symmetric (1D) version of one of these systems, which is explicitly
given in the Appendix A.

The source terms in (\ref{balance}) take into account the nonlinear
part Einstein's equations.
We have used an operator
splitting approach by considering separately the source driven evolution
\begin{equation}
\partial_t {\bbox u} = {\bbox S}({\bbox u})	\label{sources}
\end{equation}
and the transport process
\begin{equation}
\partial_t {\bbox u} + \partial_\rho {\bbox F}({\bbox u}) =
0.\label{transport}
\end{equation}
These different processes are then combined (Strang splitting
\cite{Recipes}) to obtain a second order accurate discretization of
the full equation (\ref{balance}).

In order to analyze the importance of the source terms in the overall
evolution, we have implemented both a standard second order
Runge-Kutta along with a sophisticated high order Bulirsch-Stoer
method \cite{Recipes}. We have found that the accurate modeling of
the transport step was far more significant than the source treatment
in the overall evolution.

We have used a second order upwind method
(see Appendix B) to deal with the
transport step. Let us note that the hyperbolicity of the system
(\ref{transport}) allows one to express it as two uncoupled subsystems
with the structure of (the first order form of) the wave
equation\cite{PRL92}. This is useful when implementing the upwind method
because one can easily construct the linear combinations of the original
variables which propagate along light rays going in the forward or backward
directions (characteristic variables, see the Appendix).

The boundary conditions at the inner boundary are then to be imposed
on the forward propagating combinations only. This is done by allowing
for the inversion symmetry (\ref{inversion}) at the throat (we will
give more details in Sec.~ 5). Conversely, the outer boundary
conditions will affect only to the backward propagating combinations.

\section{FIXED GRID RESULTS.}
We have performed our computations with an evenly
spaced grid of 200 points ranging from 1 to 40 Schwarzschild radii,
thus obtaining a resolution that can be already tested in 3-D
codes\cite{3dcode}, thus allowing future comparison of results.
The accuracy was monitored by computing the mass function (\ref{Bondi}),
which will keep its constant value $M(t,\rho)=m$ only if the constraint
equations are preserved. We have found that this provides an
extremely sensitive error test.

The results presented in Fig.~\ref{fig:mass} actually show the
mass function, as computed from the numerically evolved quantities
($m=2$ in our case). The time values correspond to the proper time of
the outermost evolving point. Errors are big around the horizon position,
which is computed from (\ref{horizon}) and marked with a cross, in spite
of the huge dynamical range one gets at the black hole throat
as one approaches the singularity, as it is clearly shown
in Fig.~\ref{fig:metric} for the radial metric component.

This apparent paradox is explained by the collapse of
the speed of light in the inner zone which locally freezes the
evolution there, as it is clearly shown in Fig.~\ref{fig:light}.
This behavior amounts to the well known collapse of the lapse
which is generic to singularity avoiding coordinate systems, like
harmonic or maximal slicing. It allows us to deal with the huge
dynamical range in the metric components near the singularity (see
Fig.~\ref{fig:metric}), but it has a perverse side effect: the
appearance of steep gradients near the horizon which are actually
the main source of numerical errors.

One can of course try to reduce these errors by increasing either the
grid resolution or the accuracy of the numerical method. We have
rather preferred, in keeping with Ref.~\cite{Seidel}, to
avoid the gradients by evolving just the exterior part of the black
hole spacetime (a moving boundary problem) with the moving grid
techniques sketched in the following section.

\section{A MOVING GRID APPROACH.}
Let us introduce a new radial function $r$, which is related to the
previous one $\rho$ in a time dependent way:
\begin{equation}
\rho = f(t,r),						\label{change}
\end{equation}
so that a grid of fixed points with respect to the new variable $r$
will be moving with respect to the original one, attached to
$\rho$. The first order derivatives of (\ref{change})
\begin{equation}
V \equiv \partial_t f, \;\;\;\;\;\; \Delta \equiv \partial_r f
\label{derivatives}
\end{equation}
can be interpreted as the relative speed and dilation factor,
respectively, as computed by an observer attached to the moving grid.

The structure of the set of conservation laws (\ref{transport}) does
not change when written in terms of the new radial function
\begin{equation}
\partial_t\:(\Delta\:{\bbox u}) + \partial_r\:
    [{\bbox F}({\bbox u})-V\:{\bbox u}]=0,
\label{moving}
\end{equation}
where we have transformed the quantities ${\bbox u}$ in a scalar
way. The new characteristic matrix keeps the same set of eigenvectors
as the original one, so that hyperbolicity is preserved. The new
characteristic speeds are obtained from the old ones simply by
subtracting the grid speed $V$ and then dividing by the dilation
factor $\Delta$. Notice that the change of variables (\ref{change})
introduces just one new degree of freedom, the dilation factor being
obviously related to the grid speed:
\begin{equation}
\partial_t \Delta - \partial_r V = 0.		\label{dilation}
\end{equation}

Even the static choice $V=0$ gives us the possibility of choosing a
nontrivial dilation factor to deal with numerical grids which are not
evenly spaced with respect to the original variable $\rho$. We
actually used it to obtain the fixed grid results of the previous
section.  The reason has to do with the treatment of the inner
boundary: the image under (\ref{inversion}) of an evenly spaced set of
points is no longer evenly spaced. Equations (\ref{moving}) with $V=0$
provided an elegant way to preserve the overall accuracy at the inner
boundary \cite{inner} without running into stability problems.

The next simplest case is the linear one, where the dilation factor
depends on the time coordinate only. This leads to a linear speed
profile which can be determined by the boundary values. One can apply
this to the black hole problem by demanding the grid speed to coincide
with the local speed of light (characteristic speed) at the
horizon. In that way, one is placing the inner boundary at the horizon
(instead of at the throat), which will remain attached to a specific
node of the moving grid. The profile can be fully determined by
demanding the grid speed to vanish at (or near to) the external
boundary. We have found this characteristic boundary approach very
convenient to avoid the steep
gradient zone behind the horizon wile keeping a high degree of
accuracy in the black hole exterior, as we show in the following
section.

\section{MOVING GRID RESULTS.}
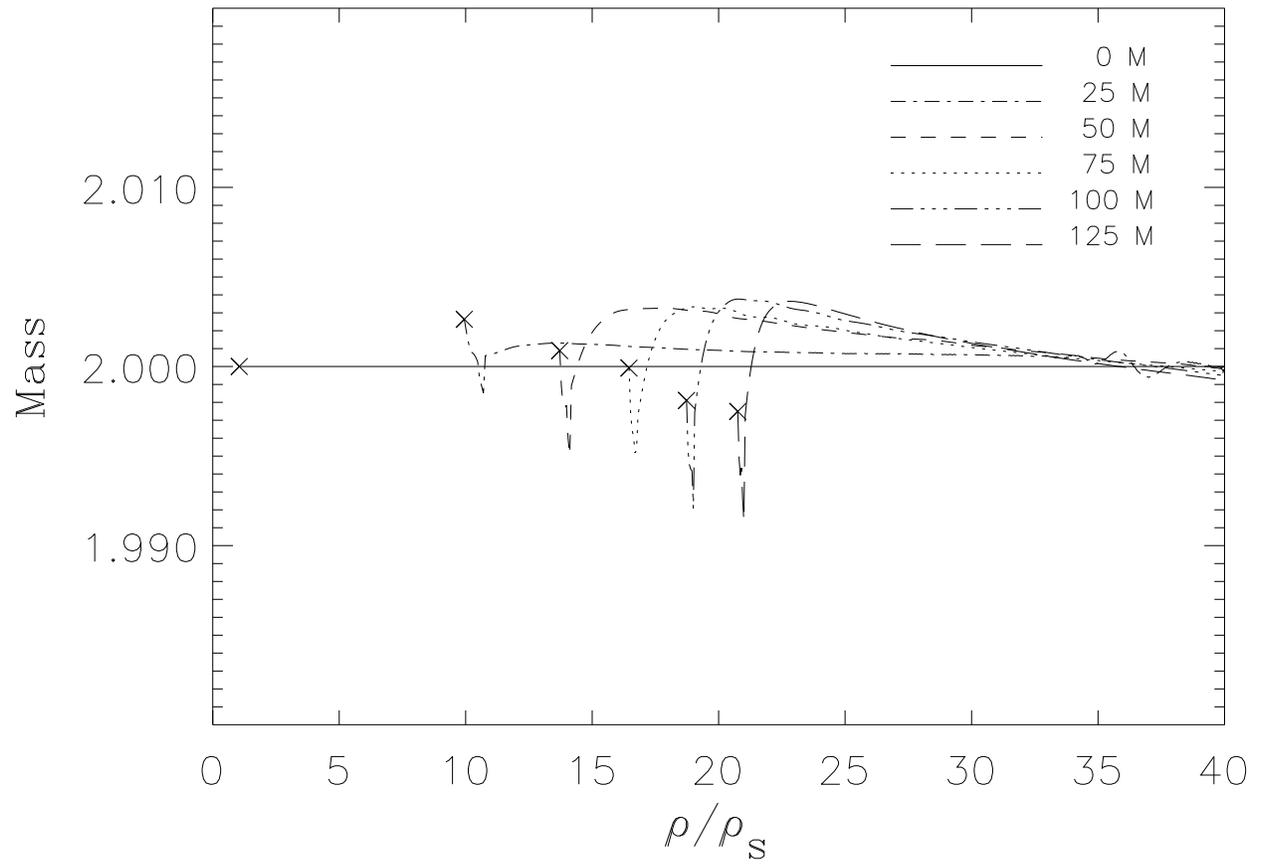
\begin{figure}
\caption{Same as Fig.~1 for the moving grid case. The maximum error is
now less than 0.5\%. Note that the scale has been enlarged by one
order of magnitude.}
\label{fig:massmoving}
\end{figure}

We have redone with the moving grid the same computations presented in
Sec.~4, under the same initial conditions. The linear relationship
between the fixed and moving grid coordinates allows us to plot our
moving grid results in terms of the original variable $\rho$, allowing
a direct comparison with Sec.~4. In that sense,
Fig.~\ref{fig:massmoving} is to be compared with the previous
Fig.~\ref{fig:mass}. The successive plots start now at the horizon
position because the inner part is no longer computed. Note that the
error in the mass has decreased drastically: this confirms
that the larger errors in Fig.~\ref{fig:mass} were due to the steep
gradient zone behind the horizon.

Note that we are now placing 200 evenly spaced points in the region
between the horizon and the outer boundary. The evolution can then be
pursued until this region gets very small, leading to an extremely low
dilation factor (extremely high characteristic speed) which freezes
the evolution. In this moving grid approach, the
lifetime of a numerical black hole is just the time it takes the
horizon to arrive to the grid outer boundary \cite{infall}.

We have also tested other initial values for the lapse function,
different from (\ref{lapse1}). We found that the fixed grid code
crashed in some cases, due to large errors, while the moving grid one
kept its low error profile in every case.  We conclude that the moving
grid approach is both a more robust and more accurate way to deal with
the 1D black hole problem, significantly improving the accuracy of the
results in every case.

\section{RELATED APPROACHES.}
A similar approach, which would be more appealing to relativists, is
to consider the transformation (\ref{change}) as a change of spacetime
coordinates, not just affecting the grid motion, but also the metric
quantities. The line element (\ref{metric}) would then transform into:
\begin{equation}
ds^2=-\alpha^2(t,r)\:dt^2+X^2(t,r)\:(\Delta\:dr+V\:dt)^2+Y^2(t,r)\:d\Omega^2
\label{new metric} \end{equation}
so that a radial shift vector appears:
\begin{equation}
\beta^r=V/\Delta					\label{shift}
\end{equation}
and the radial metric component gets multiplied by the dilation factor.

The new degree of freedom can now be used in a different way, by
imposing a fixed form for the radial metric function and adding the
shift $\beta$ to the list of the variables \cite{Seidel}. The constant
value of the radial metric function ensures that the use of a
non-uniform dilation factor does not lead to a premature freezing of
the evolution. But the main price to pay is that the structure of the
original system (\ref{transport}) is modified at the risk of losing
hyperbolicity and, with that, the possibility of applying standard
numerical algorithms. We are currently studying the structure of the
general 3+1 (or ADM) evolution system in order to see whether or not
hyperbolicity can actually be preserved when introducing shift vectors
and/or when using other algebraic gauge conditions.

There are of course other adaptive grid approaches, based on mesh
refinement algorithms \cite{Berger}, which have been successfully
applied to the 1D scalar field case \cite{Matthew}. Such powerful
methods include sophisticated bookkeeping routines to manage the grid,
putting more resolution just where it is needed, without modifying the
equations in any way. The application of these methods to the generic
3D case is a big challenge to the present day computing and
programming capabilities.

\section{DISCUSSION AND OUTLOOK.}

We are aware that many of the difficulties one would encounter in
evolving really dynamic spacetimes are avoided in the Schwarzschild
test.  As a first example, even in the 1D case, let us remember that a
black hole can be formed in the collapse of a supermassive star. The
horizon forms at a given instant in the evolution and new horizons can
appear later. The evolution of the inner boundary will be then
piecewise continuous, jumping every time that a new outermost horizon
appears.

As stated in section 2, our code contains an horizon finding routine,
based on the apparent horizon local definition (\ref{horizon}), which
can detect when and where a new horizon does appear. This routine can
also detect whether the horizon is actually ahead of the inner
boundary due to cumulative numerical errors in computing the light
speed there. We have also implemented another routine which, when the
difference exceeds two grid zones, makes the inner boundary to jump
from its previous location (the first grid point) to the actual
horizon position, discarding the grid points between the old and new
positions and recomputing the speed profile accordingly. We have found
that a few of these jumps do not compromise the stability of the code
and therefore we believe that the discontinuous horizon motion in 1D
dynamical spacetimes can be dealt with by using the same technique.

One would encounter other difficulties when trying to extend this
formalism to the multidimensional black hole case. For instance, the
horizon can expand in an anisotropic way so that the resulting speed
profile can lead to a highly distorted numerical mesh. To solve this
problem, one can make use of the relationship (\ref{shift}) between
the grid speed and the shift vector. One could demand that the
velocity field satisfies either the ''minimal strain'' or ''minimal
distortion''\cite{Smarr} or a similar condition. The corresponding set
of elliptic equations should be solved by using the light speed
components as inner boundary values.

But possibly the main difficulty in going to the 3D case is that the
horizon finding routine can not be a straightforward generalization of
the 1D one. The multidimensional analogous of (\ref{horizon}) contains
explicitly the unit normal to the horizon surface (something one does
not know a priori). This is the heel of Achilles of 3D adaptive grid
black hole codes. We recently heard about promising progress on that
subject\cite{recent1,recent2}, and it seems that it will help in
solving such difficulty in a near future.

\acknowledgments
We want to thank Dr.~Edward Seidel for useful discussions and
suggestions.  This work is supported by the Direcci\'on General para
la Investigaci\'on Cient\'{\i}fica y T\'ecnica of Spain under project
PB91-0335. J.M. acknowledges a Fellowship (P.F.P.I.) from Ministerio
de Educaci\'on y Ciencia of Spain
and NSF grant PHY/ASC93-18152 (Arpa supplemented).

\appendix
\section{EVOLUTION EQUATIONS.}
The set of metric variables to evolve is
$${\bbox u} \equiv\left(C, \Gamma_r,
  g^{rr},g^{\theta\theta},
  q^r{}_r,q^\theta{}_\theta,
  D^r{}_r,D^\theta{}_\theta
\right)$$
where $C \equiv {\alpha\sqrt{g^{rr}}}$ is the local speed of light and
we have introduced the shortcuts $D^r{}_r \equiv \partial_r
ln(g^{rr})$, $D^\theta{}_\theta \equiv \partial_r
ln(g^{\theta\theta})$, $\Gamma_r \equiv D^\theta{}_\theta-\frac{1}{2}
D^r{}_r - L_r$, and $L_r \equiv \partial_r ln(\alpha)$.

With this notation, the vacuum Einstein Evolution Equations in
spherical symmetry can be written in first order form as follows:
\begin{mathletters}
\begin{equation}
   \partial_t C = -C^2 q^\theta{}_\theta  \;,
\end{equation}
\begin{equation}
   \partial_t \Gamma_r = C\left[2 q^\theta{}_\theta L_r +
           (q^\theta{}_\theta - q^r{}_r) D^\theta{}_\theta\right] \;,
\end{equation}
\begin{equation}
   \partial_t g^{rr} = C g^{rr} q^r{}_r                      \;,
\end{equation}
\begin{equation}
   \partial_t g^{\theta\theta} = C g^{\theta\theta} q^\theta{}_\theta  \;,
\end{equation}
\begin{eqnarray}
   \partial_t \left(q^r{}_r+q\right)
 - \partial_r \left[C(D^r{}_r+2\Gamma_r-2L_r)\right]
	= C \left[(q^\theta{}_\theta)^2-(D^\theta{}_\theta)^2\right] \;,
\end{eqnarray}
\begin{eqnarray}
   \partial_t \left(q^\theta{}_\theta+q\right)
 - \partial_r \left[C(D^\theta{}_\theta-2L_r)\right]
= C \left[ q^\theta{}_\theta (2q^\theta{}_\theta - q^r{}_r)
 +2 \frac{g^{\theta\theta}}{g^{rr}}-(D^\theta{}_\theta)^2
 +2  L_rD^\theta{}_\theta\right],
\end{eqnarray}
\begin{equation}
   \partial_t \left(D^r{}_r-2 L_r\right) =
   \partial_r \left[C(q^r{}_r+q)\right]  \;,
\end{equation}
\begin{equation}
   \partial_t \left(D^\theta{}_\theta-2 L_r \right) =
   \partial_r \left[C(q^\theta{}_\theta+q)\right] \;,
\end{equation}
where we have also noted $q \equiv q^i{}_i = q^r{}_r+2q^\theta{}_\theta$.
\end{mathletters}

The nontrivial characteristic quantities are
\begin{eqnarray}
w^r_{\pm} =  C(D^r{}_r+2\Gamma_r-2L_r) \pm (q^r{}_r+q)\;, \;\;\;\;\;\;
w^\theta_{\pm} =  C(D^\theta{}_\theta-2L_r) \pm (q^\theta{}_\theta+q)\;,
\end{eqnarray}
with characteristic speeds $\pm C$.

\section{DISCRETIZATION OF THE TRANSPORT PROCESS.}

The finite difference version of the transport equation
\begin{equation}
\partial_t {\bbox u} + \partial_\rho {\bbox F}({\bbox u}) = 0
\end{equation}
is constructed in two steps.

In the first step, one computes all the variables at the cell interfaces
(points $i\pm\frac{1}{2}$) at an intermediate time level $n+\frac{1}{2}$.
The resulting predictions
\begin{equation}
{\bbox u}^{n+1/2}_{i\pm 1/2}
\end{equation}
are then used to compute the Fluxes
\begin{equation}
{\bbox F}^{n+1/2}_{i\pm 1/2}  =  {\bbox F}({\bbox u}^{n+1/2}_{i\pm 1/2}),
\end{equation}
so that, in the second step, the next time level is computed in an explicit
flux-conservative way:
\begin{equation}
{\bbox u}^{n+1}_i  = {\bbox u}^n_i -\mbox{$\frac{dt}{dx}$}
                 \left({\bbox F}^{n+1/2}_{i+1/2}
                      -{\bbox F}^{n+1/2}_{i-1/2}\right).
\end{equation}

Note that the stability of this explicit scheme implies that the time
interval of every step is limited by the causality condition ensuring
that the characteristic speeds are lower than one grid zone per
timestep.

Allowing for the fact that this second step is centered both in space
and in time, first order accuracy in the interface values leads to
second order accuracy in the final result, because the leading error
terms cancel when subtracting the Fluxes between the right and left
interfaces.

In order to obtain the interface values, we first compute the forward and
backward first order predictions for every quantity,
\begin{equation}
{\bbox u}^{FW}_{i-1/2}  = \mbox{$\frac{3}{2}$}{\bbox u}^n_{i-1}
-\mbox{$\frac{1}{2}$}{\bbox u}^n_{i-2}-\mbox{$\frac{1}{2}\frac{dt}{dx}$}
                 \left({\bbox F}^n_{i-1} -{\bbox F}^{n}_{i-2}\right),
\end{equation}
\begin{equation}
{\bbox u}^{BW}_{i-1/2}  = \mbox{$\frac{3}{2}$}{\bbox u}^n_i
-\mbox{$\frac{1}{2}$}{\bbox u}^n_{i+1}-\mbox{$\frac{1}{2}\frac{dt}{dx}$}
                 \left({\bbox F}^n_{i+1} -{\bbox F}^{n}_i\right).
\end{equation}

Note that, when dealing with shocks or large gradients, these
one-sided predictions are known to produce spurious oscillations: the
predicted values at the $i-\frac{1}{2}$ interface may get out of the
interval defined by the values at the grid points $i$ and $i+1$. This
is anomalous in the sense that we are modeling a transport process and
the causality condition ensures that nothing coming from outside of
this interval can reach the interface at the intermediate time
level. We detect this spurious behaviour when one of the anomalous
sequences ($u^n_{i}$, $u^n_{i+1}$, $u^{FW}_{i+1/2}$) or
($u^{BW}_{i+1/2}$, $u^n_{i}$, $u^n_{i+1}$) is monotonic. In these
cases, we take $u^{FW}_{i+1/2}=u^n_{i+1}$ or $u^{BW}_{i+1/2}=u^n_{i}$,
respectively, to make sure that our predictions do not get out of
range.

We perform then a local transformation at every interface from our
original quantities ${\bbox u}$ to the set characteristic variables
${\bbox w}$ which are given in the preceding Appendix. The
hyperbolicity of our transport equations ensures that this is a one to
one invertible transformation. The interface values for each
characteristic component $w$ are taken to be either the forward or
backward prediction, depending on the sign of the corresponding
characteristic speed (positive or negative, respectively). The
interface values of the original quantities ${\bbox u}$ are finally
recovered by inverting the transformation.

\begin{figure}
\caption{Plots of the mass function (fixed grid case), which should be
equal to 2 everywhere. The maximum error is around 15\%.  The
successive horizon positions are marked with a cross. Note that the
different plots coincide in the inner part, where evolution is
freezing due to the light speed collapse (see Fig.~3).}
\label{fig:mass}
\end{figure}

\begin{figure}
\caption{Sucessive plots of the time evolution of the logarithm of the
metric component $g_{rr}$.  The crosses show the successive positions
of the black hole horizon.  Steep gradients appear just behind the
horizon.  The resulting dynamical range rises up to eighty orders of
magnitude.}
\label{fig:metric}
\end{figure}

\begin{figure}
\caption{Plots of the coordinate light speed, showing a sudden
collapse of the inner part (the black hole interior), due to the
singularity avoidance of the harmonic slicing.}
\label{fig:light}
\end{figure}

\begin{references}
\bibitem{apparent}	The use of the term 'black hole horizon' is not
			misleading here because, in the Schwarzschild case,
			the event horizon and the apparent horizon do coincide.
\bibitem{inner} Note however that the resulting dilation factor is not smooth
		(just continuous) at the inner boundary, so that second order
		accuracy does not make much sense there.
\bibitem{infall} We can put this in the converse way: the numerical observers
		are falling into the static black hole and the limit is the
		infall time of the last observer.
\end{references}
\end{document}